\documentclass[prd,aps,showpacs,nofootinbib]{revtex4}
\usepackage{amssymb}
\usepackage{graphicx}
\usepackage{amsmath}

\newcommand{\be}{\begin{equation}}
\newcommand{\ee}{\end{equation}}
\newcommand{\bq}{\begin{eqnarray}}
\newcommand{\eq}{\end{eqnarray}}

\begin{document}

\title{Unruh effect as foundation of universal gravitation within the cosmological scenario}
\author{Cl\'audio Nassif and A. C. Amaro de Faria Jr.\\
(e-mail: cncruz777@yahoo.com.br, antoniocarlos@ieav.cta.br*)}

\altaffiliation{{\bf CBPF}: Centro Brasileiro de Pesquisas F\'isicas, Rua Dr.Xavier Sigaud, 150, Urca, CEP:22.290-180,
 Rio de Janeiro-RJ, Brazil.\\
{\bf **UTFPR-GP}: Federal Technological University of Paran\'a, Avenida Professora Laura Pacheco Bastos, 800, Bairro Industrial, 
CEP: 85.053-510, Guarapuava-PR, Brazil.\\
{\bf **IEAv}: Institute of Advanced Studies, Rodovia dos Tamoios, Km 099, CEP: 12.220-000, S\~ao Jos\'e dos Campos-SP, Brazil.}

\date{\today}

\begin{abstract}
We aim to build a simple model of a gas with temperature ($T$) in thermal equilibrium with a black-body that plays the role of the 
adiabatically expanding universe, so that each particle of such a gas mimics a kind of ``particle'' (quantum) of dark energy, which is
inside a very small area of space so-called Planck area ($l_p^{2}$), that is the minimum area of the whole space-time represented
by a huge spherical surface with area $4\pi r_u^2$, $r_u$ being the Hubble radius. So we should realize that such spherical surface 
is the surface of the black-body for representing the universe, whose temperature ($T$) is related to an acceleration ($a$) of a proof 
particle that experiences the own black-body radiation according to the Unruh effect. Thus, by using this model, we derive the law of
universal gravitation, which leads us to understand the anti-gravity in the cosmological scenario and also estimate the tiny order of 
magnitude of the cosmological constant in agreement with the observational data. 
\end{abstract}

\maketitle

\section{\label{sec:level1} Introduction}

The origin of gravity is still a puzzle in spite of the success of General Relativity (GR) in considering a curved space-time
due to the presence of a mass and its well-known relevant implications. Such puzzle arises from the quantum aspect of gravitation, which 
occurs in the so-called Planck length scale $l_p\sim 10^{-35}$m related to Planck energy scale $E_p\sim 10^{19}$GeV with
Planck temperature $T_p\sim 10^{32}$K, which represents a quantum singularity where the gravitational field equation of GR (Einstein's
equation) fails. So, in the early universe with temperature $T_p\sim 10^{32}$K, vacuum fluctuations prevailed in a scenario of
quantum-gravity, which is not still well-understood. Currently, the most widely accepted explanation for their origin is in the context of 
cosmic inflation. 

The statistical properties of the primordial fluctuations can be inferred from observations of anisotropies in the cosmic microwave
background and from measurements of the distribution of matter. Since the fluctuations are believed to arise from inflation, such
measurements can also set constraints on parameters within inflationary theory.

Another puzzle that should be deeper explored is the thermodynamical aspect of gravitation. Recent investigations have shown that 
gravitational field equations in a wide class of models can be interpreted with a thermodynamical origin. This idea, originally due to 
Sakharov\cite{1}, has different forms of implementation \cite{1}\cite{2}\cite{3}\cite{4}\cite{5}\cite{6}\cite{7}\cite{8}\cite{9}\cite{10}
\cite{11}\cite{12}\cite{13}\cite{14}\cite{15}\cite{16} and also other approaches \cite{17}\cite{18}\cite{19}.

\section{\label{sec:level1} Gravity and cosmological antigravity from Unruh effect}

In the early universe with Planck temperature $T_p$, space-time was confined to Planck area $l_p^{2}$. So imagine such an extremely small 
area $l_p^{2}$, which earlier has expanded dramatically (inflationary model) and then continued to expand more softly
to reach the current area $4\pi r_u^2$, $r_u(\sim 10^{26}$m) being the Hubble radius. Thus, in this holographic model, the expanding 
universe 3-D works like the surface 2-D of a spherical ``balloon'' with area $4\pi r^2$ for a given age and temperature. In other words, 
we say that such holographic model is considering a spherical surface 2-D for representing the own ``tissue'' of space-time 4-D that
expands, since a quantum space 2-D given in the Planck scale ($l_P^2$) leads to a continuum space 3-D seen in the macro-scale, 
according to this holographic model. 

According to this model, let us admit the universe as being a gas of $N$ particles in thermal equilibrium with the ``particles'' on 
the surface of the cavity of the black-body (spherical surface) that plays the role of a background frame for representing the ``tissue'' 
of space-time. Furthermore, since the ``cavity 2-D of the black-body'' is the own space-time (universe), we should consider that each particle
of the gas represents a proof particle that mimics and has a correspondent virtual ``particle'' inside a very small area of the dark
cavity, that is the Planck area ($l_p^{2}$). In doing this, we conclude that the spherical surface, that is the universe in this
holographic model, can be divided in $N$ areas of Planck $l_p^2$, so that we write

\begin{equation}  
 N=\frac{4\pi r^2}{l_p^2}, 
\end{equation}
where $N$ is the number of ``particles of such gas'' and $l_p$ is the Planck length, namely: 

\begin{equation}
 l_p=\sqrt{\frac{G\hbar}{c^3}}
\end{equation}

Since each ``particle of such gas'' over the spherical surface is expanding radially in only one direction $r$, let us write
the thermal energy of the gas as follows: 

\begin{equation}
 E=mc^2=\frac{1}{2}NK_BT 
\end{equation}

Now, by inserting Eq.(2) into Eq.(1) and after into Eq.(3), we obtain 

\begin{equation}
 E=mc^2=\left(\frac{2\pi c^3 K_B}{G\hbar}\right)Tr^2, 
\end{equation}
where $r$ is the radius of the sphere. 

Here we must stress that such ``gas'' is not a true gas where its particles can move randomly over a $2$-dimensional surface 
(spherical surface). Actually, it is a imitation of a gas considered for only one degree of freedom (radial direction $r$), 
so that each ``particle'' of this ``gas'' is connected to a quantum of space-time given by the Planck area $l_p^2$, which is radially 
expanding (Fig.1). 

\begin{figure}
\begin{center}
\includegraphics[scale=0.6]{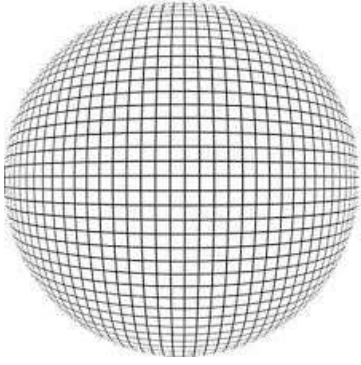}
\end{center}
\caption{The surface of the sphere (expanding universe) is formed by a number $N$ of minimal areas (Planck areas), each one of them
being connected to a dark infinitesimal energy working like a ``particle'' of an exotic ``gas'', whose total mass is $m_{dark}$. 
A baryonic proof particle of a certain gas in thermal equilibrium with the sphere with temperature $T$, being placed over its surface 
experiences an acceleration according to the Unruh effect. Such acceleration has anti-gravitational origin, which is related 
to the cosmological constant.}
\end{figure}

Now let us consider a proof particle (e.g: a particle of a true gas) over the spherical surface with temperature $T$ [Eq.(4)], such that
this particle is in thermal equilibrium with the sphere. Then, in this case, as this particle has a radial acceleration ($a=a(r)$), 
the Unruh effect\cite{20} states that such a particle experiences a thermal bath with the same temperature $T$ of the sphere [Eq.(4)], 
namely: 

\begin{equation}
 T=\frac{\hbar a}{2\pi K_B c} 
\end{equation}

Due to the thermal equilibrium between the accelerated particle (proof particle) and the spherical surface with temperature $T$, we
can insert Eq.(5) into Eq.(4) and thus we get 

\begin{equation}
 mc^2=\left(\frac{2\pi c^3 K_B}{G\hbar}\right)\left(\frac{\hbar a}{2\pi K_B c}\right)r^2, 
\end{equation}
from where we finally obtain 

\begin{equation}
 a=\frac{Gm}{r^2}, 
\end{equation}
where $m$ is the mass of the spherical surface, $r$ is its radius and $a$ is the radial acceleration (acceleration of gravity)
of a proof particle over the surface. 

Eq.(7) is the own law of universal gravitation and it is interesting to call attention to the fact that such law was obtained in a cosmological
scenario, where different concepts like black-body, background radiation (Unruh effect) and an expanding exotic ``gas'' with mass $m$ for
representing the space-time quantized in Planck areas are put in together, all of them being essentially connected by a quantum gravity
(Planck scale), at least in the context given by Eq.(1). 

Although Eq.(7) is the law of universal gravitation, which can be applied for any spherical body with mass $m$ that 
generates a gravitational force of attraction over a proof particle, Eq.(7) was obtained in a scenario of quantum-gravity [Eq.(1)] connected to a vacuum
energy, i.e., a dark mass $m$ of a virtual ``gas'' that fills the whole space-time. In this sense, since such ``gas'' is expansive, 
leading to the expansion of the spherical surface (Fig.1), we realize that the acceleration obtained in Eq.(7) is an anti-gravitational 
acceleration, which occurs only in the cosmological scenario where we consider the whole universe, so that the mass $m$ in Eq.(7) 
is a kind of dark mass that fills the whole universe represented by the spherical surface of radius $r$ for a given age of
the expanding universe. 

In short, we can say that Eq.(7) is the Newton's gravitation law applied to the expanding universe as a whole
(a spherical dark mass $m_{dark}$ with a certain radius $r$); however Eq.(7) works like an anti-gravitational law. Thus, we should 
rewrite Eq.(7), as follows: 

\begin{equation}
 a=a_{exp}=-\frac{Gm_{dark}}{r^2},
\end{equation}
where the sign ``$(-)$'' simply indicates that the acceleration is repulsive, $a_{exp}$ being the acceleration of expansion,
$r$ is the radius of the sphere (universe) and $m_{dark}$ is the dark mass related
to the dark energy $E_{dark}=m_{dark}c^2$, which is about $73$ per cent of the whole universe. So, a proof particle (luminous matter like a
gas of cosmic dust, a star or a galaxy) on the surface of such dark sphere (frontier of the expanding universe) experiences a repulsive 
acceleration that will be estimated by Eq.(8), from where we will also derive the cosmological constant. 

According to recent studies\cite{21}, the density of matter in the universe is about $3\times 10^{-30}g/cm^3$, which means that it is 
$300$ billion billion billion times less dense than water. Note that this includes the contribution of dark matter (non-baryonic matter)
plus the density of luminous matter or baryonic matter (that we see as stars and galaxies), which is about $27$ per cent of the whole 
universe, about $23$ per cent being dark matter and $4$ per cent being luminous matter. 

Now, the size of the observable universe is about 14 billion light years, and using the above value of density related to luminous
and dark matter ($27$ per cent of the whole universe), we find a mass of about $3\times 10^{55}$g; however, since dark energy is
the majority of the universe ($73$ per cent), we estimate $m_{dark}$ (related to dark energy) about $8.1\times 10^{55}$g, which leads
us to conclude that the entire energy (mass) content can be roughly estimated as being the own dark energy $m_{dark}$ of the order of
$10^{56}$g. Then, by making $m_{dark}\sim 10^{53}$Kg, $r=r_u\sim 10^{26}$m and $G\sim 10^{-10}Nm^2/kg^2$ in Eq.(8), we obtain the following 
acceleration of anti-gravity for a proof particle, namely $a\sim 10^{-9}m/s^2$, which is an expected result because the acceleration 
of expansion of the universe is too small. 

We know that the cosmological ``constant'' $\Lambda$ is related to the acceleration of expansion of the universe. Actually $\Lambda$
is the acceleration rate per unit of distance $r$ ($s^{-2}$ in SI units), so that we have $\Lambda=da/dr$. In doing this, from Eq.(8),
we find

 \begin{equation}
 \Lambda=\frac{da}{dr}=\frac{2Gm_{dark}}{r^3},    
 \end{equation}
where we realize that $\Lambda$ has not been constant over the cosmological time, since it depends on $r$; however the so-called
cosmological constant is given for the Hubble's radius $r_u\sim 10^{26}$m. 

 Now, from Eq.(9), we obtain the well-known (current) cosmological constant, namely: 

\begin{equation}
\Lambda=\Lambda_{o}=\left[\frac{da}{dr}\right]_{r=r_u}=\frac{2Gm_{dark}}{r_u^3}
\end{equation}

As we have $G\sim 10^{-10}Nm^2/kg^2$, $m_{dark}\sim 10^{53}$kg and $r_u\sim 10^{26}$m, thus, from Eq.(10), we finally obtain
$\Lambda_{o}\sim 10^{-35}s^{-2}$, which is just the order of magnitude of the cosmological constant in agreement with the observational
results\cite{22}\cite{23}\cite{24}\cite{25}\cite{26}\cite{27}\cite{28}

We would like to call attention to the fact that the very high values ​​obtained for the cosmological constant and the vacuum energy 
density by means of Quantum Field Theory (QFT) for describing the quantum vacuum have a discrepancy of about 120 orders of magnitude
beyond their observational values. This puzzle is well-known as the {\it Cosmological Constant Problem} 
(http://www.worldscientific.com/worldscinet/ijmpd?journalTabs=read)\cite{29}, introduced by Zeldovich and well discussed by 
Weinberg and Padmanabhan\cite{30}.

\end{document}